\documentclass{ws-rv9x6}
\usepackage{subfigure}   
\usepackage{ws-rv-thm}   
\usepackage{ws-rv-har}
\usepackage{aas_macros}

\makeindex

\begin{document}

\chapter[The exoplanet hosts]{Twenty-five years of exoplanet discoveries:\\
The exoplanet hosts\label{ch1}}

\author[B. Rojas-Ayala]{B\'arbara Rojas-Ayala}

\address{Instituto de Alta Investigación, Universidad de Tarapacá,\\
Casilla 7D, Arica, Chile,\\ brojasayala@uta.cl}

\begin{abstract}
For centuries, humanity wondered if there were other worlds like ours in the Universe. For about a quarter of a century, we have known that planetary systems exist around other stars, and more than 3800 exoplanetary systems have been discovered so far. However, the large majority of the exoplanets remain invisible to us since we usually infer their presence by their effect on their star. The chapter is devoted to stellar hosts and their characteristics, emphasizing their description by discovery method and links between the properties of the host stars and their planets. The star-planet connection is vital to constrain the theories on the formation and evolution of planetary systems, including our own.
\end{abstract}

\smalltoc

\tableofcontents

\body

\section{The relevance of the properties of the planet hosts}

The discovery of new worlds has been inevitably linked to studying the stars for the past twenty-five years. The most successful detection methods for planets (radial velocity and transit techniques) measure the effect on the star caused by the exoplanet, not the exoplanet itself. Hence, the properties of those new worlds are derived from the observables and the properties of their host stars. For example, radial velocity semi-amplitude $K$,
\begin{equation}
K \approx \left( \frac{2\pi G}{P M_\star^2} \right)^\frac{1}{3} \frac{M_{\rm
    planet} \sin{i}}{\sqrt{1-e^2} },
\end{equation}
and transit depth $\delta_{\rm tra}$,
\begin{equation}
  \delta_{\rm tra} \approx  \left(\frac{R_{\rm planet}}{R_\star}\right)^2~\left(1 - \frac{I_{\rm planet}(t_{\rm tra})}{I_\star}\right),
\end{equation}
\\
are observables from the radial velocity and transit techniques. To obtain the properties of the bulk properties of the exoplanets,$R_{\rm planet}$ and $M_{\rm planet}$, we need to know the bulk properties of the star, $R_{\rm \star}$ and $M_{\rm \star}$, respectively.  
\\
\\
The properties of host stars are needed because:
\begin{itemize}
    \item we want to know how planet formation works and what determines their evolution, 
    \item we make target selection for exoplanet searches (e.g., input catalogs for space-based missions)
    \item we want to ensure that what we are measuring is due to a planet around the star and not a false positive (e.g., activity, rotation).
\end{itemize}

\begin{figure}[ht]
\centerline{\includegraphics[width=7cm]{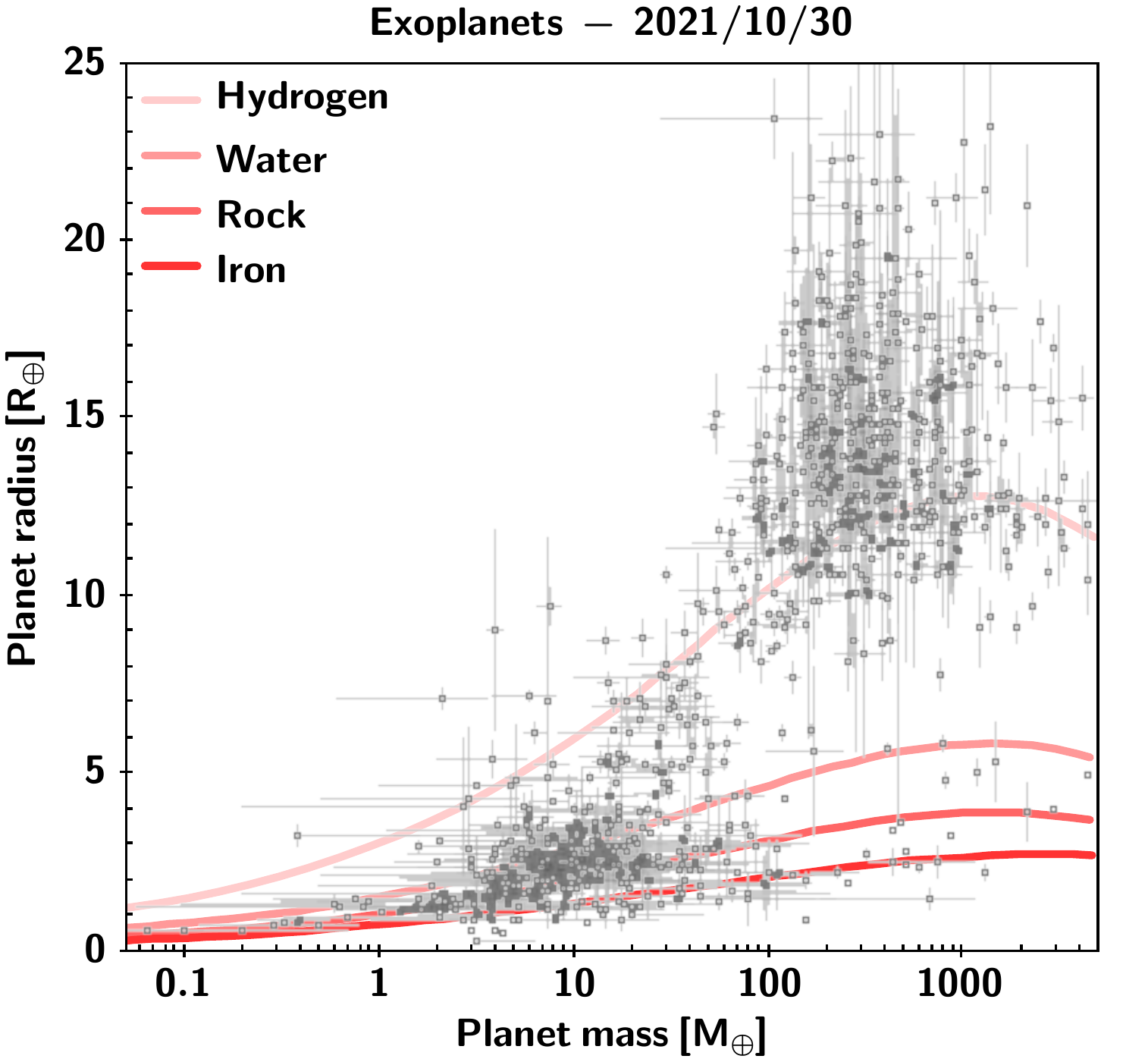}}
\caption{ M-R relation: observations vs. theoretical data. The circles correspond to the planets up to October 30$^{th}$ 2021, while the solid lines represent mass-radius relations for different planetary compositions.} \label{plotmassradius}
\end{figure}

\fref{plotmassradius} shows exoplanets with mass and radius estimates in the {\it NASA Exoplanet Archive} up to October 30$^{th}$ 2021, along with mass-radius relationships for planets with pure iron, rock  (Mg$_2$SiO$_4$) and water ice compositions from \citet{2007ApJ...669.1279S} and pure hydrogen composition from \citet{2007ApJ...659.1661F}. The Mass-Radius diagram for the discovered planets shows us the diversity of worlds being found and makes plain evident the necessity to improve the precision of their mass and radius to constrain their composition. Over the past years, stable spectrographs and space telescopes have provided exquisite data to measure the observables precisely, but it is not enough for some hosts because the uncertainties on their masses and sizes are pretty significant. Therefore, the exact exoplanet flavor will depend on how well we know the bulk properties of the host star. 

The most fundamental property of a star is its mass. However, masses are not easy to directly measure for most stars. We can get precise masses for stars in binary systems, and if they are eclipsing binaries, we can get their accurate sizes. Stellar sizes can be obtained from interferometry if the star is relatively bright and we know how far the star is from us (e.g., from their Hipparcos/GAIA parallaxes). Asteroseismology is a powerful tool for insights into the stellar interior and obtaining the stellar mass, radius, and ages with high precision if the star pulsates. In particular, all of the above becomes more challenging for the low-mass stars due to their low luminosities and lack of detected pulsations in photometric and spectroscopic data up-to-date. Since the large majority of planet hosts do not satisfy the conditions above, the exoplanet community has relied mainly on the atmospheric stellar parameters ($T_{\it eff}$, $[M/H]$ and $log\: g$) estimates of their bulk properties. For example,  you can get a precise estimate of the  $T_{\it eff}$ of the star from high-resolution spectra, and if you know its parallax and luminosity, you can derive its radius. Then, from the estimate of the star's surface gravity, you can obtain its mass. Stellar evolution models have been beneficial in deriving masses and sizes of stars from atmospheric stellar parameters.

\begin{figure}[ht]
\centerline{\includegraphics[width=7cm]{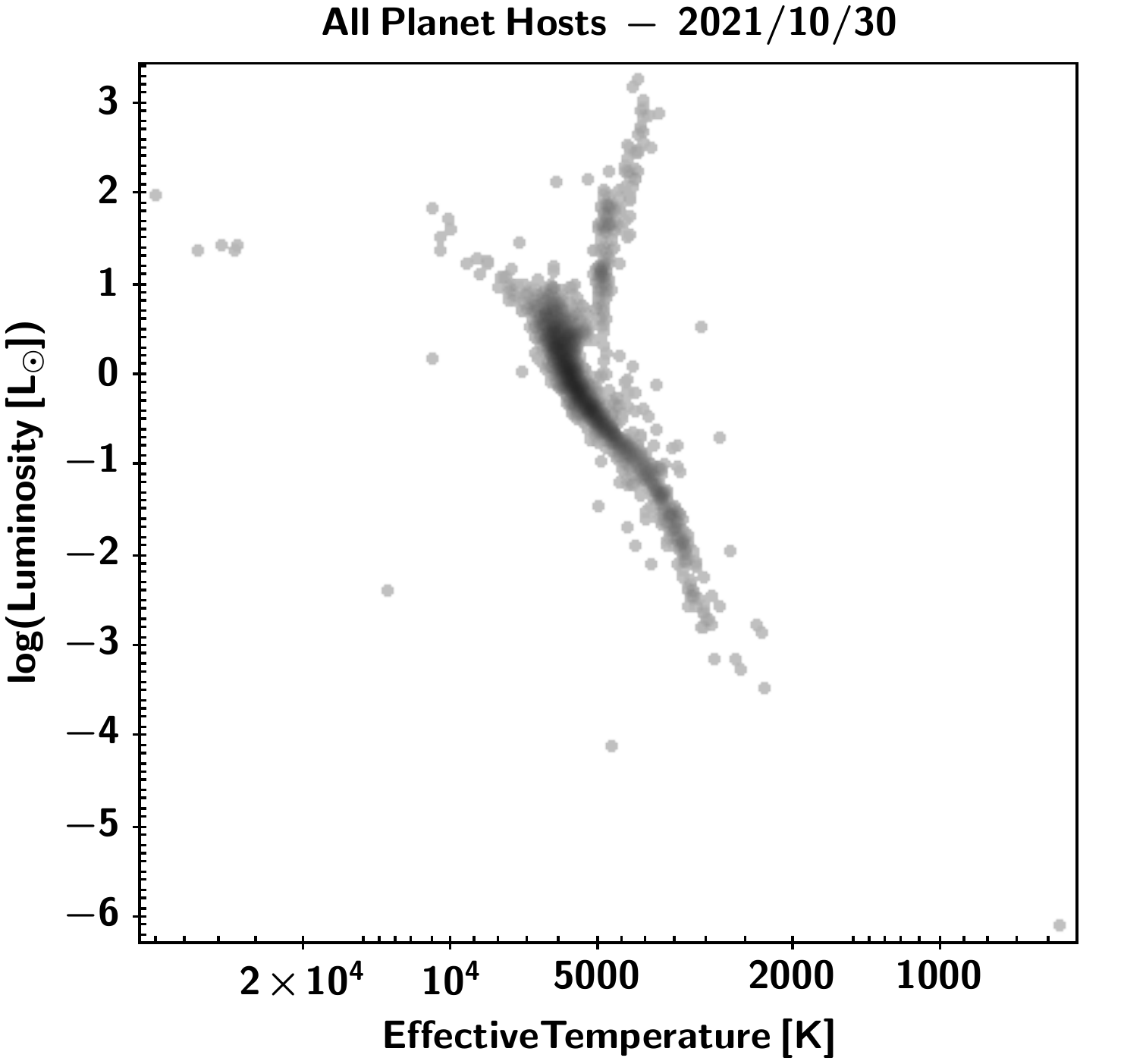}}
\caption{The planet hosts with luminosity and effective temperature estimates from the {\it NASA Exoplanet Archive} up to October 30$^{th}$ 2021.} \label{plotHR}
\end{figure}

\section{Characteristics of the confirmed stellar hosts up to October 2021}

According to the {\it NASA Exoplanet Archive}, up to October 30$^{th}$ 2021, there were 4451 planets in 3378 systems. The {\it NASA Exoplanet Archive} is an astronomical catalog and data service that collects and cross-correlates relevant information on exoplanetary systems such as stellar, exoplanet, and discovery/characterization data. Thus, it serves as a census of exoplanetary systems constantly being updated and available to all. Unfortunately, not all the confirmed planet hosts in the NASA Exoplanet Archive are fully characterized, meaning they are missing estimates of effective temperature, metallicity, surface gravity, mass, radius, and/or luminosity. In fact, the Hertzsprung-Russell diagram constructed with the data available up to October  30$^{th}$ in the archive shows only 3247 hosts out of the 3378 systems. About 4$\%$ of the hosts do not have effective temperature and/or luminosity estimations.  \fref{plotHR} shows a couple of peculiar hosts, such as white dwarfs and hot subdwarfs, since most hosts are main sequence, subgiant, and giant stars. The lack of specific stellar parameters for the planet host is somewhat related to the detection technique involved in the exoplanet discovery.

\begin{figure}[ht]
\centerline{\includegraphics[width=7cm]{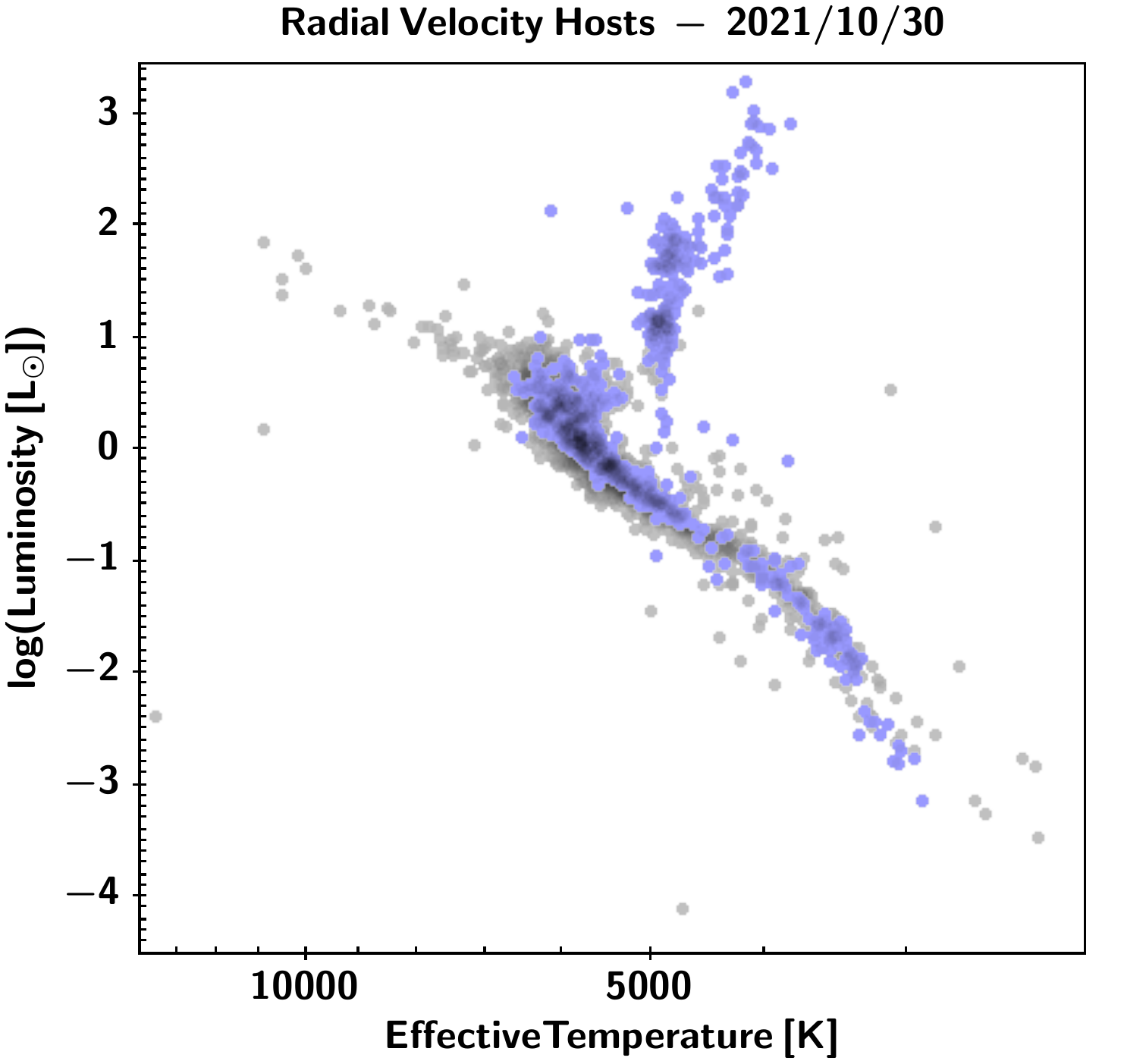}}
\caption{The stars with planets found with the radial velocity technique in the {\it NASA Exoplanet Archive} up to October 30$^{th}$ 2021 are shown in purple.} \label{plotRVHR}
\end{figure}

\subsection{Radial velocity hosts}
The locations of the hosts discovered by the radial velocity (RV) technique are shown in \fref{plotRVHR}. The RV technique makes it easier to find planets around main-sequence (GK) stars and (sub)giant stars (bright/slow) since relatively bright stars provide high signal-to-noise observations. It is harder to find planets around F and earlier stars because of the lack of absorption lines to analyze the data correctly. It is also more challenging to find planets around young stars due to their activity and variability. Stellar activity can be a problematic signal to remove from the data. Spots, plages, convection, and pulsations can induce RV signals to reach amplitudes larger than a planet's signal. However, RV observations with near-infrared spectrographs have facilitated the discovery of planets around M dwarfs and young stars.

\begin{figure}[ht]
\centerline{\includegraphics[width=7cm]{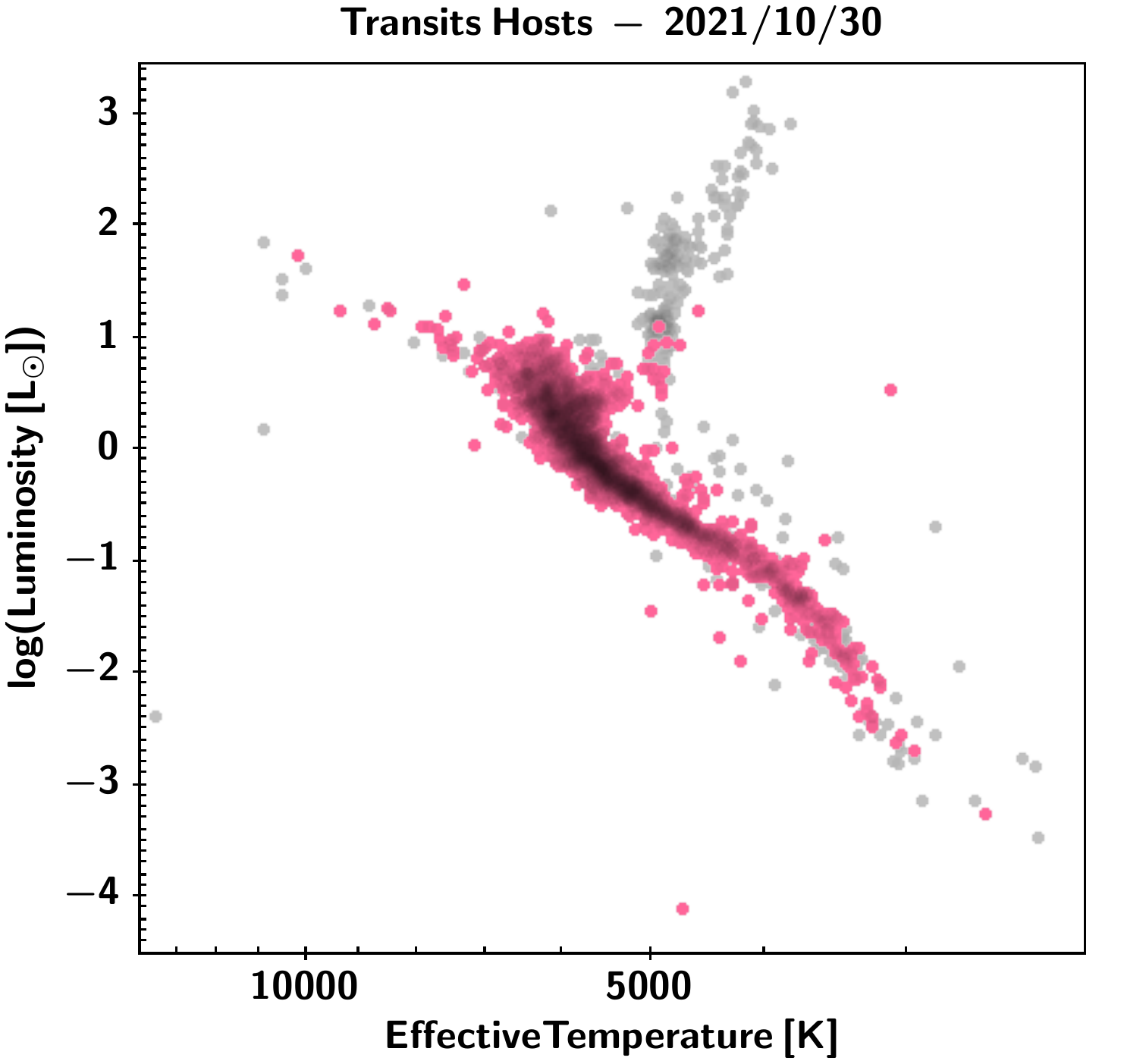}}
\caption{The stars with planets found with the transits technique in the {\it NASA Exoplanet Archive} up to October 30$^{th}$ 2021 are shown in pink.} \label{plotTranHR}
\end{figure}

\subsection{Transit hosts}
The locations of the hosts discovered by the transit technique are shown in \fref{plotTranHR}. The transit technique works best in bright, small, and inactive stars. Small stars are an advantage for the transit technique because the drop in luminosity is proportional to the ratio between the size of the planet and the star. It is easier to find planets around GK dwarfs, and bright M dwarfs since relatively bright stars provide high signal-to-noise observations. It is harder to find planets around evolved stars since they are too large. It is also harder to find planets around young stars because of their variability. Most of the transit hosts in \fref{plotTranHR} were discovered by the Kepler and K2 missions.

\begin{figure}[ht]
\centerline{\includegraphics[width=7cm]{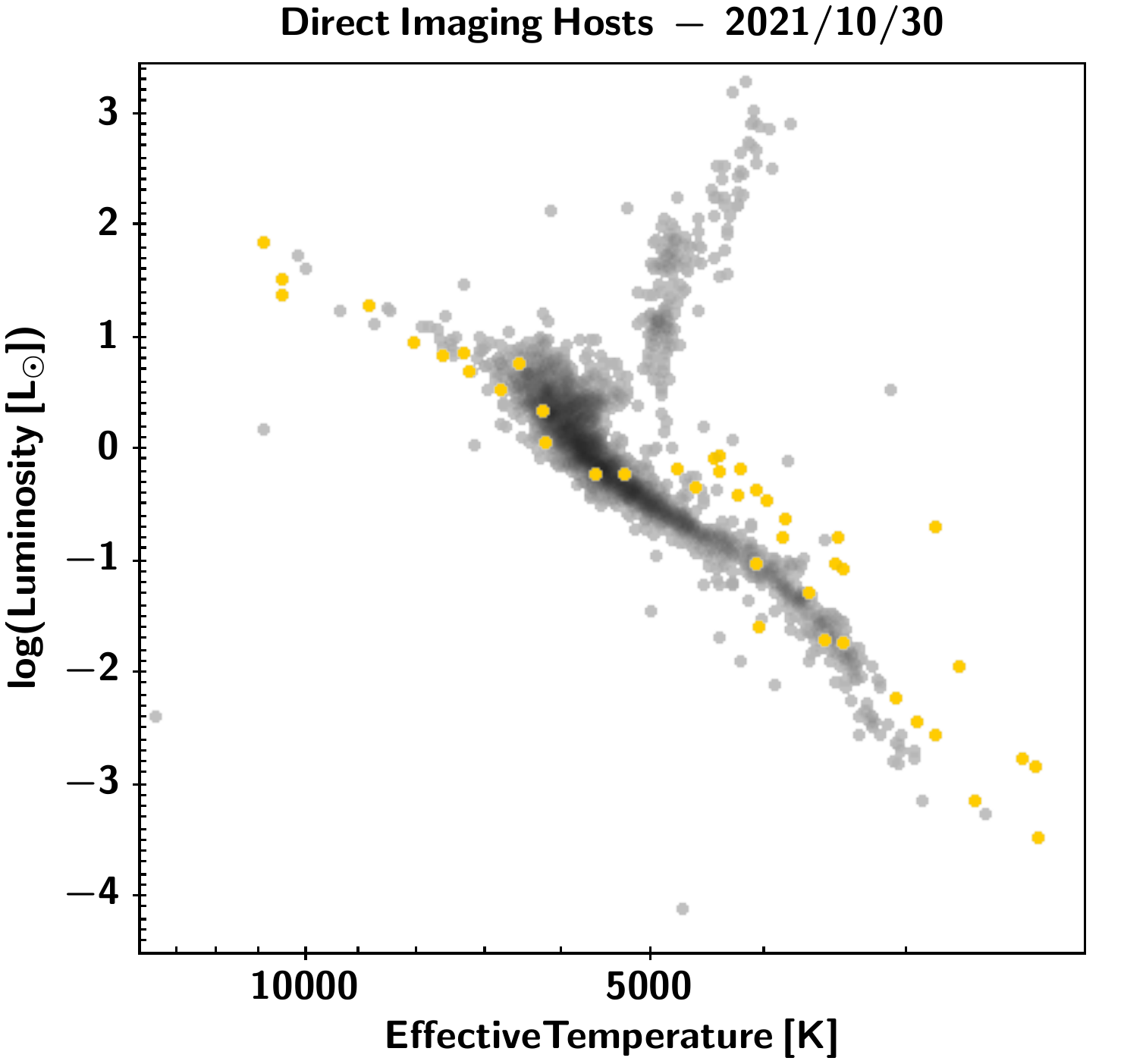}}
\caption{The stars with planets found with the direct imaging technique in the {\it NASA Exoplanet Archive} up to October 30$^{th}$ 2021 are shown in yellow.} \label{plotImHR}
\end{figure}

\subsection{Direct imaging hosts}
The locations in the Hertzsprung-Russell diagram of the hosts discovered by the direct imaging technique are shown in \fref{plotImHR}. This technique performs best around nearby and young stars. It is easier to find planets around young A stars and nearby young associations because the worlds are still contracting and, therefore, are brighter than in systems where the host has reached the main-sequence branch. On the other hand, it is harder to find planets around evolved and main-sequence stars because of the luminosity contrast between the host and the exoplanet.

\begin{figure}[ht]
\centerline{\includegraphics[width=7cm]{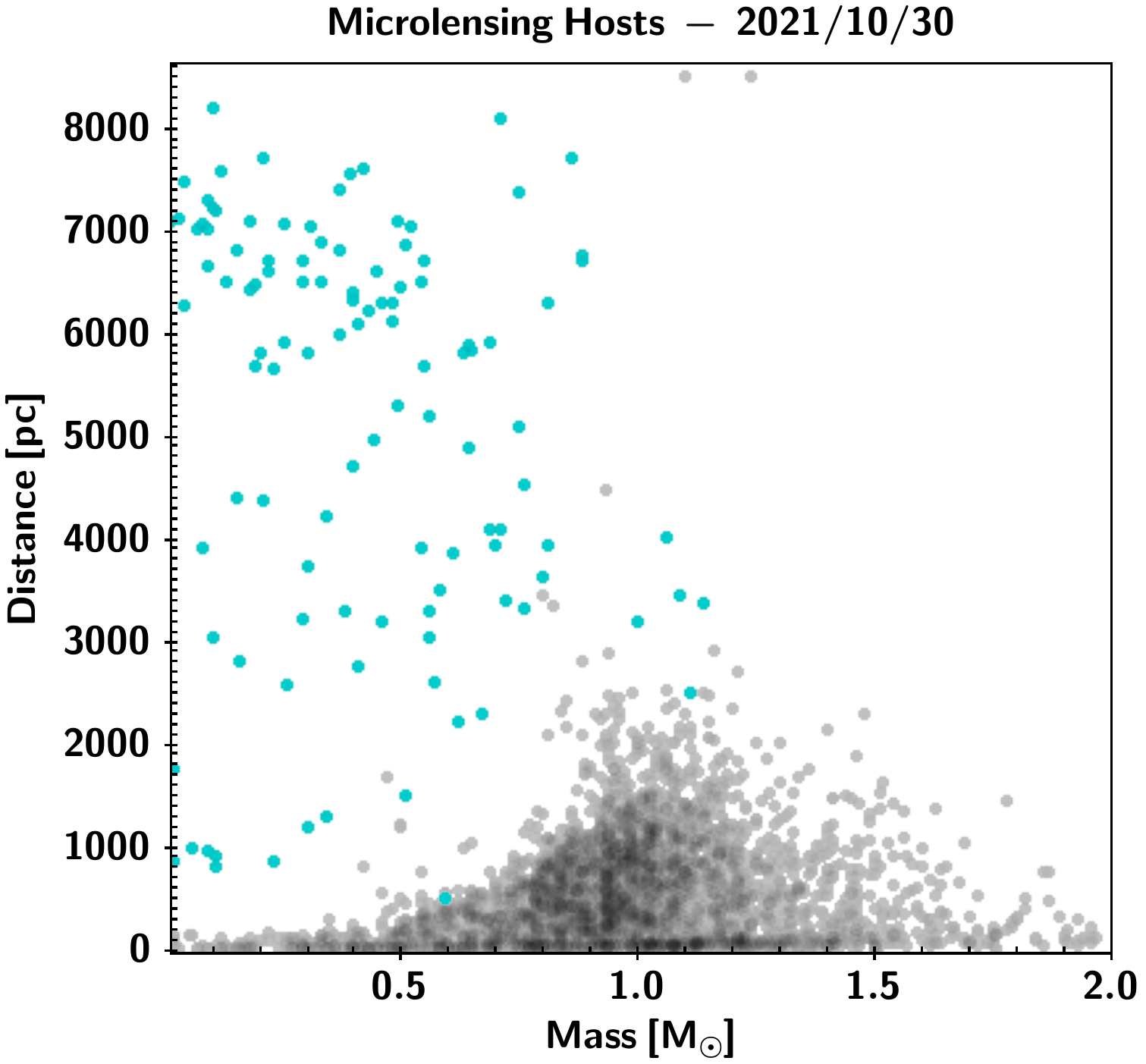}}
\caption{ The stars with planets found with the microlensing technique in the {\it NASA Exoplanet Archive} up to October 30$^{th}$ 2021 are shown in green.} \label{plotMicroHR}
\end{figure}

\subsection{Microlensing Hosts}
The microlensing technique detects the effect of an unseen planetary system on the light emitted by a distant star. The host star and planets act as lenses, and the distant star gets magnified. This technique performs best in stars in front of dense stellar regions (e.g., galactic bulge). It is easier to find planets around M dwarfs since they are the most abundant type of star; it is harder to find planets in nearby stars. The hosts are difficult to characterize since they remain unseen or cannot be resolved. Microlensing hosts, therefore, are part of the stars that do not show up in the Hertzsprung-Russell diagram in \fref{plotHR}. Stellar mass and distance estimates are a result of the fitting of the magnification curve. All hosts have masses less than 1.3 solar masses, as shown in \fref{plotMicroHR}. Effective temperatures for the star can be estimated from its mass, assuming that it is a main-sequence star.

\begin{figure}[ht]
\centerline{\includegraphics[width=\paperwidth]{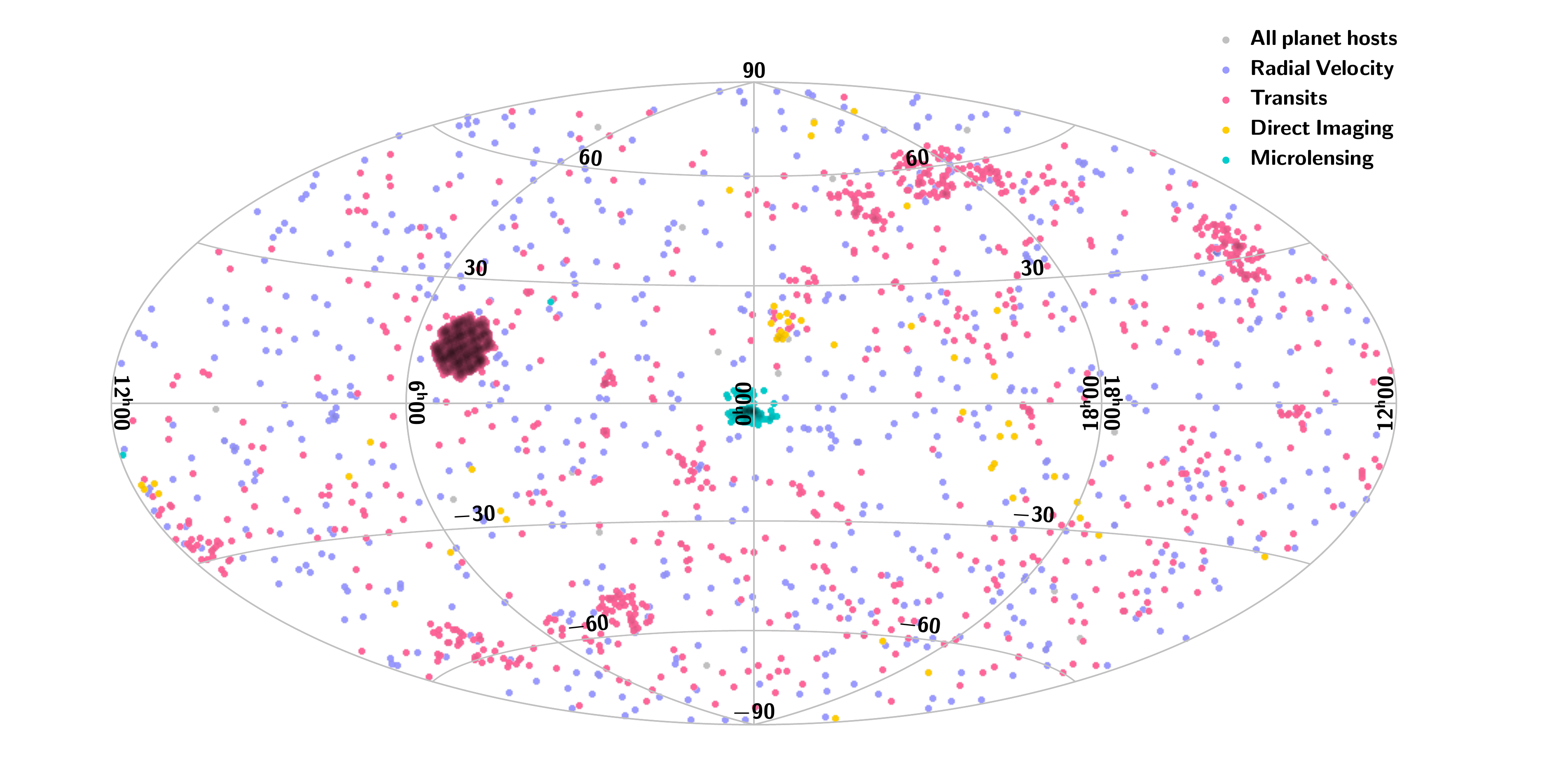}}.  
\caption{The two-dimensional projection of the positions of all planet hosts in the {\it NASA Exoplanet Archive} up to October 30$^{th}$ 2021, color-coded by the discovery technique.} \label{plotsky}
\end{figure}

\begin{figure}[ht]
\centerline{
  \subfigure[5 pc]
     {\includegraphics[width=2.3in]{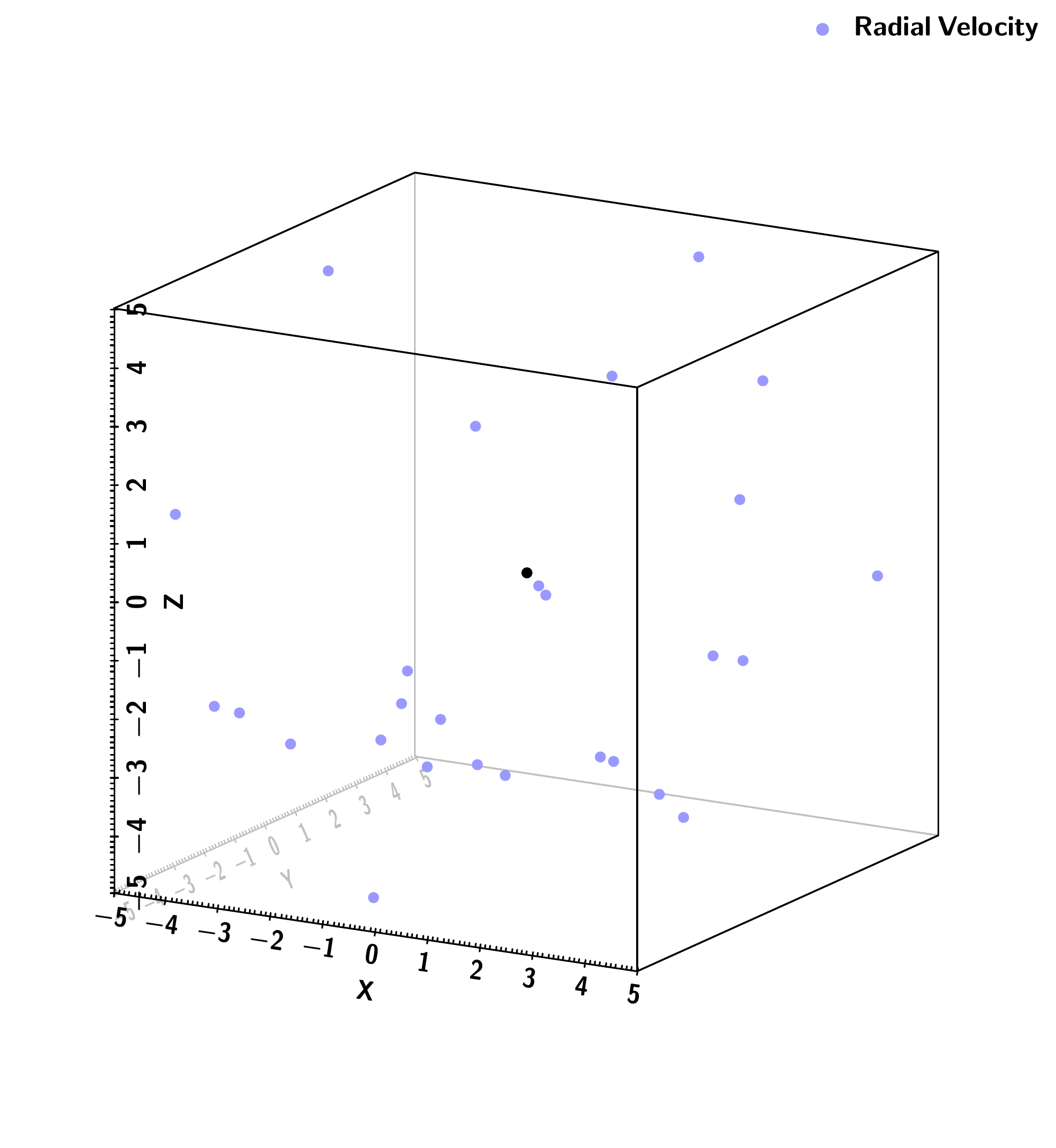}\label{plot5pc}}
  \subfigure[20 pc]
     {\includegraphics[width=2.3in]{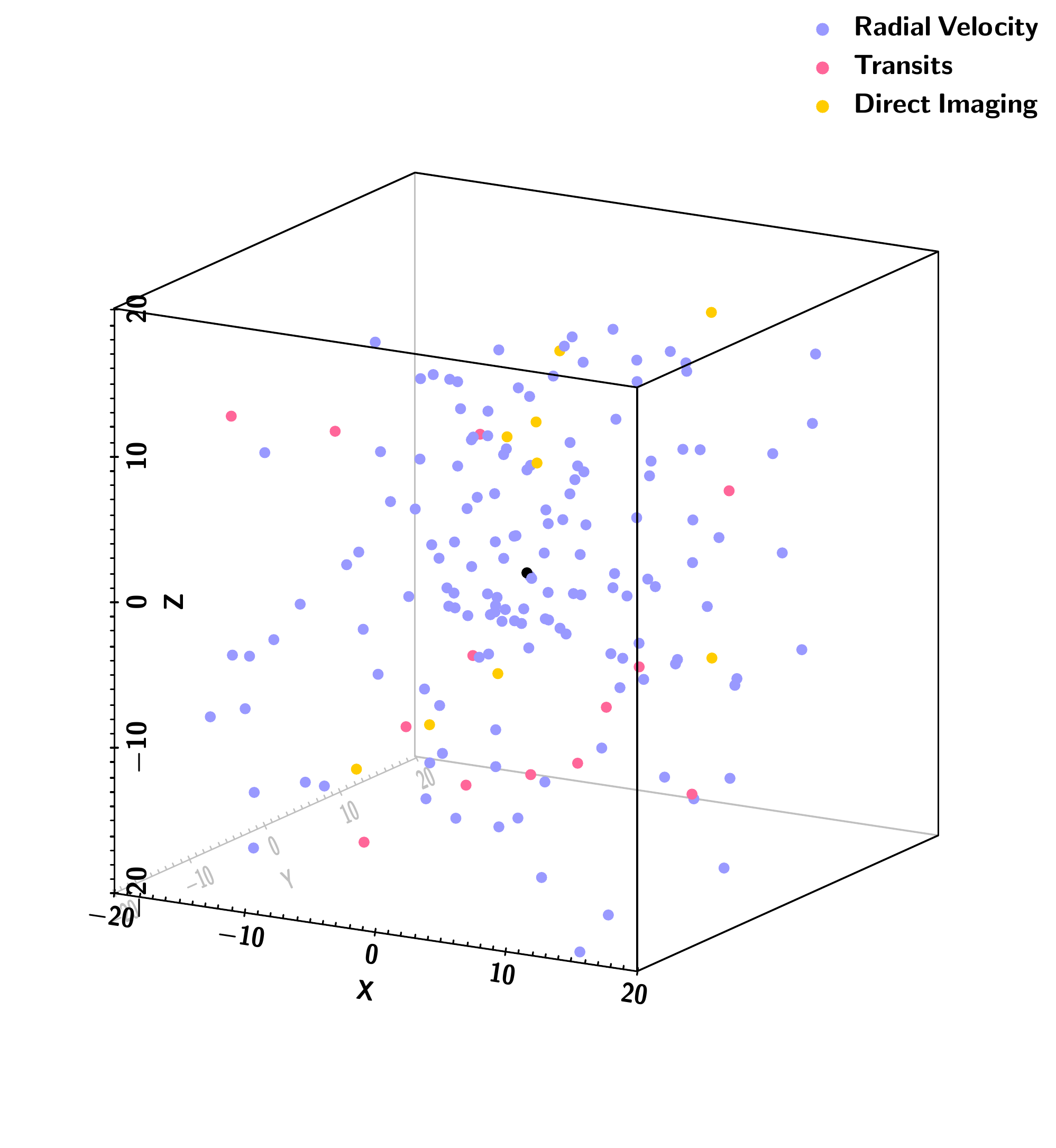}\label{plot20pc}}
     }
\centerline{
  \subfigure[500 pc]
     {\includegraphics[width=2.3in]{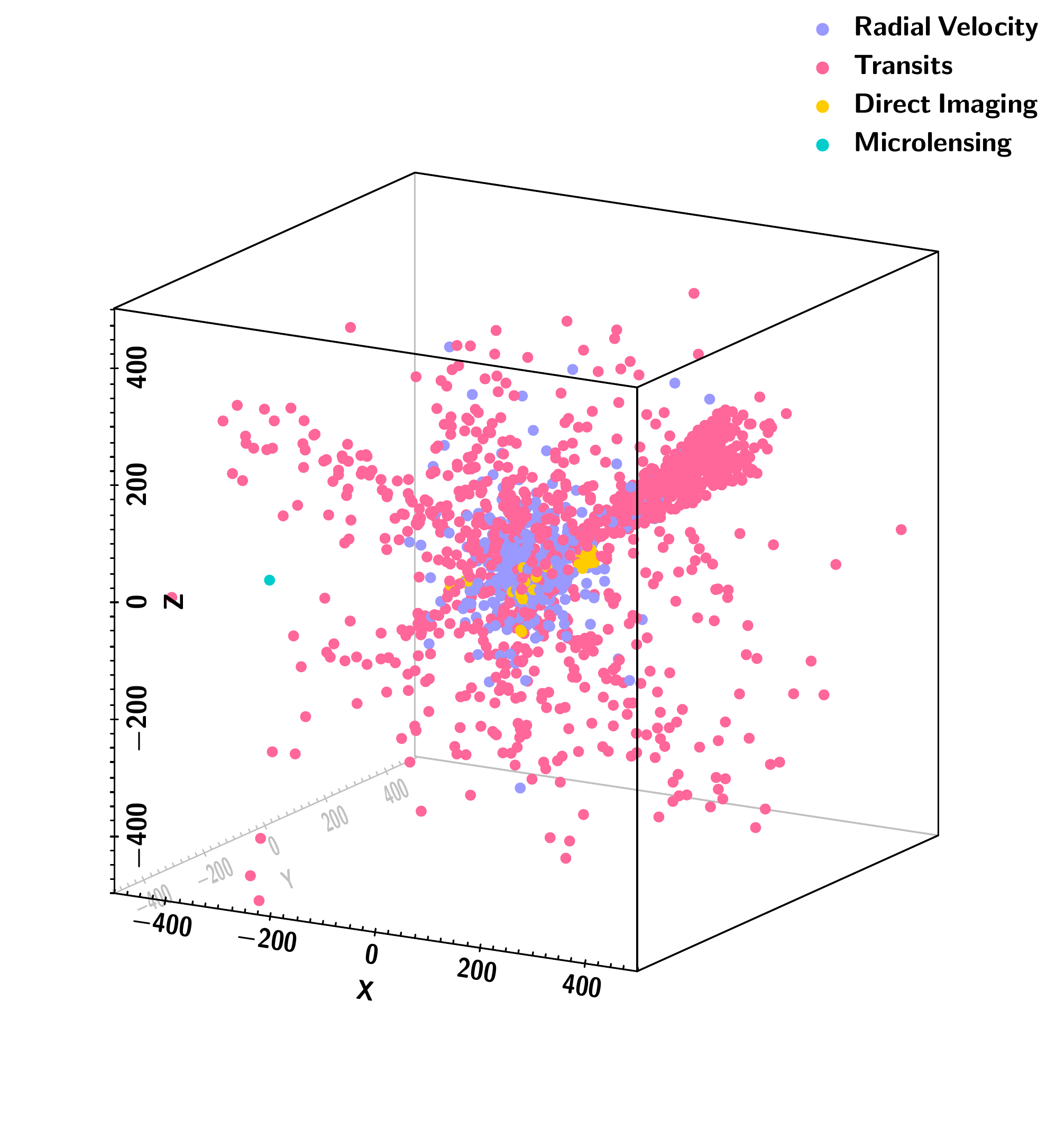}\label{plot500pc}}
  \subfigure[All systems]
     {\includegraphics[width=2.3in]{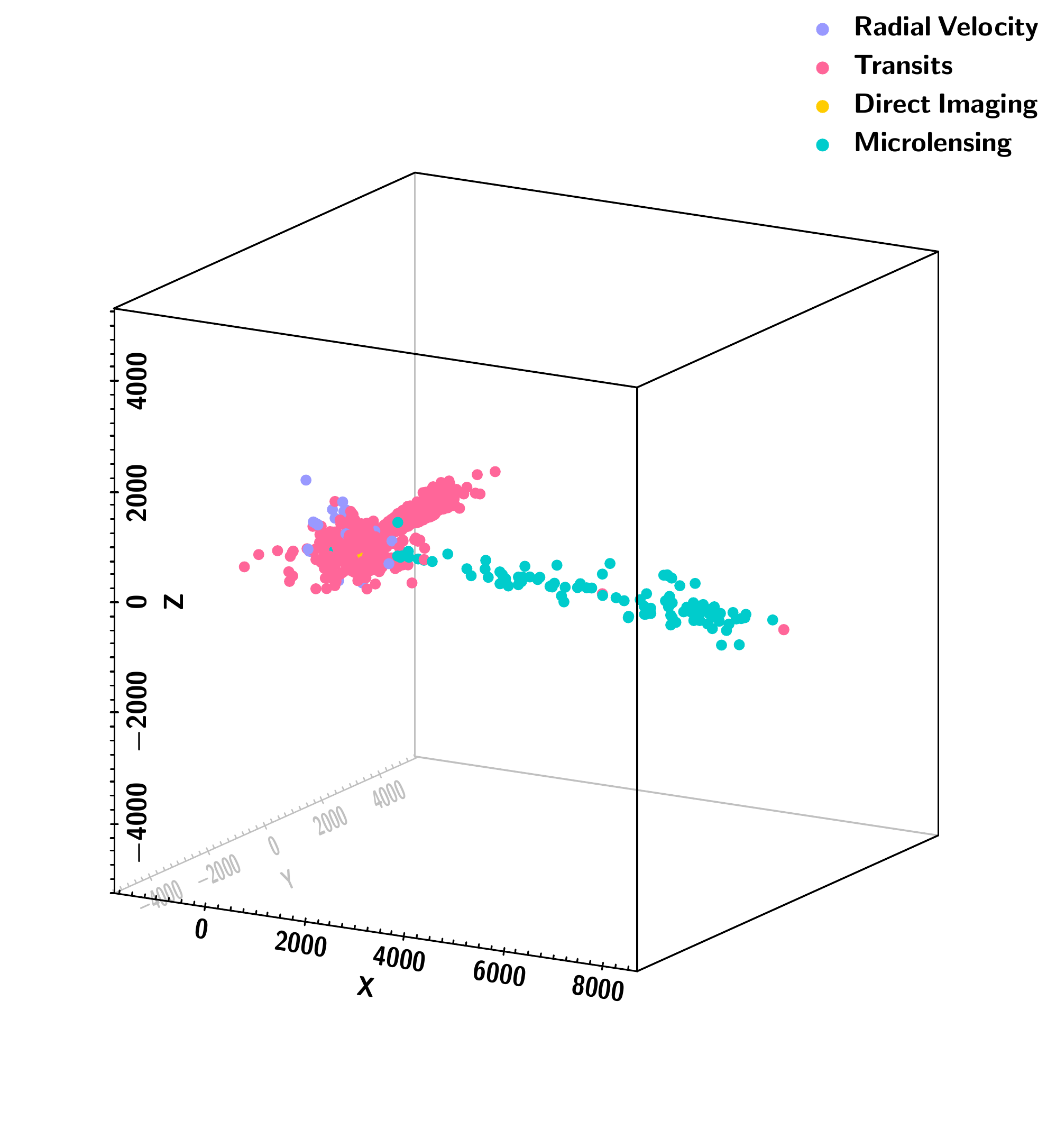}\label{plotallpc}}
     }
\caption{The three-dimensional positions of all planet hosts in the {\it NASA Exoplanet Archive} up to October 30$^{th}$ 2021, color-coded by the discovery technique.}\label{plot3d}
\end{figure}

\subsection{The sky distribution of the planet hosts}
The techniques cover different regions of the Milky Way due to the performance characteristics listed in the above sections. 
\\
In a two-dimensional representation of the sky, the radial velocity planet hosts cover roughly all radial ascensions and declinations, as seen in \fref{plotsky}. The transit hosts are also found everywhere in the projected 2D sky; however, they also bring out the Kepler and K2 fields since they cluster in those locations. The imaging hosts highlight where the young associations are found, while the majority of the microlensing hosts are located towards the bulge of the Milky Way.

In a three-dimensional representation, the limitations on the distance of the discovery methods become evident (\fref{plot3d}). At 5pc, only systems discovered by the radial velocity show up, excluding the sun (\fref{plot5pc}). At 20pc, the RV systems dominate, but transit and direct imaging systems start to show up (\fref{plot20pc}). At 100 pc, the RV systems begin to be encapsulated by the transit systems. At 500 pc, the transit systems dominate, the contribution of the Kepler mission can be clearly seen as a cone that extends from the center, and the first microlensing system appears (\fref{plot500pc}). The planetary systems found by the microlensing technique dominate at distances larger than $\sim$ 1000 pc towards the center of the Galaxy(\fref{plotallpc}).

\section{Links between the properties of the host stars and their planets}

\subsection{Occurrence rates per star type}
Each detection technique favors the discovery of planets around stars with specific characteristics. Hence, to answer how common are rocky or gaseous planets are around particular groups of stars, we need to consider the limitations of such techniques and reach a certain level of completeness for each survey. This is why occurrence rates papers started to appear roughly 10 years after discovering 51 Peg b.
Although the occurrence rates obtained from the detection methods may differ in the exact number, they are consistent in that M dwarfs have higher occurrence rates of rocky planets than FGK stars. Planet occurrence rates get updated almost every year, considering different samples that do get more complete as the searches continue. A list of articles related to planet occurrence rates can be found in the {\it NASA Exoplanet Archive} (footnote )  

\subsubsection{Examples from RV surveys} 
The RV surveys with the HARPS and CORALIE spectrographs concluded that more than 50$\%$ of the solar-type stars host at least one planet of any mass with periods up to 100 days  \citep{2011arXiv1109.2497M}. For planets with orbital periods less than 50 days and minimum masses between 3 and 30 M$_\oplus$, the occurrence rate is estimated between 15$\%$ and 27$\%$ \citep{2010Sci...330..653H, 2011arXiv1109.2497M}. \citet{2013A&A...549A.109B} estimated that about 40$\%$ of the red dwarf stars have a super-Earth orbiting in their habitable zone, and that about 12$\%$ of the red dwarfs are expected to have giant planets (100-1000 M$_\oplus$) from their M dwarf survey with HARPS. Using Lick data, \citet{2015A&A...574A.116R} concluded that the occurrence rate of giant planets in giant stars (2.7 to 5.0 M$_\odot$) is less than 1.6$\%$. \citet{2021A&A...650A..39G} estimated an occurrence rate of giant planets with periods lower than 1000 days of $\sim$ 1$\%$ for young stars.

\subsubsection{Examples from Transit surveys} 
\citet{2012ApJS..201...15H} concluded with the first results from the brightest half sample of the Kepler Mission that early M dwarfs were 7 times more likely to have a planet with an orbital period below 50 days than the hottest stars in the sample. \citet{2020MNRAS.498.2249H} estimated occurrence rates  of $\sim$ 4 or $\sim$ 8 planets per M dwarfs considering sizes between 0.5 to 4 R$_\oplus$ and periods between 0.5 and 256 days. By considering only the planets with sizes between 0.75 and 1.5 R$_\oplus$ in the habitable zone of their stars, the occurrence rate is between 0.03 to 0.40 considering orbital periods between 237 and 500 days for FGK dwarfs \citep{2019AJ....158..109H} and is 0.33 considering periods between 0.5 and 256 days for M dwarfs \citep{2020MNRAS.498.2249H}.  

\subsubsection{Examples from Direct Imaging surveys} 
If all spectral types surveyed are considered (B stars to M dwarfs), most works conclude that the occurrence rate is close to $\sim$ 1$\%$ for the most massive planets ($\sim$ 0.5-13 M$_{jup}$) at distances from few tens AU up to a few hundred AU covered by the direct imaging technique \citep[e.g.][]{2016PASP..128j2001B,2017AJ....154..129N}. If masses consistent with brown-dwarfs (up to 20 M$_{jup}$) and larger distances are considered (up to 5000 AU),  \citet{2019AJ....158..187B} estimated an occurrence rate frequency of $f={0.11}_{-0.05}^{+0.11}$ using data from several direct imaging surveys.

\subsubsection{Examples from Microlensing surveys}
 \citet{2016MNRAS.457.4089S} found that about 55$\%$ of microlensed stars host a snowline planet, and that Neptunes were about 10 times more common than Jupiter-mass planets, using OGLE, MOA, and WISE data.
 However, from twenty years of OGLE survey data, \citet{2021AcA....71....1P} found a higher occurrence rate, estimating that, on average, every microlensing star hosts at least 1 giant planet at separations from $\sim$5 AU to $\sim$15 AU.

\begin{figure}[ht]
\centerline{\includegraphics[width=8cm]{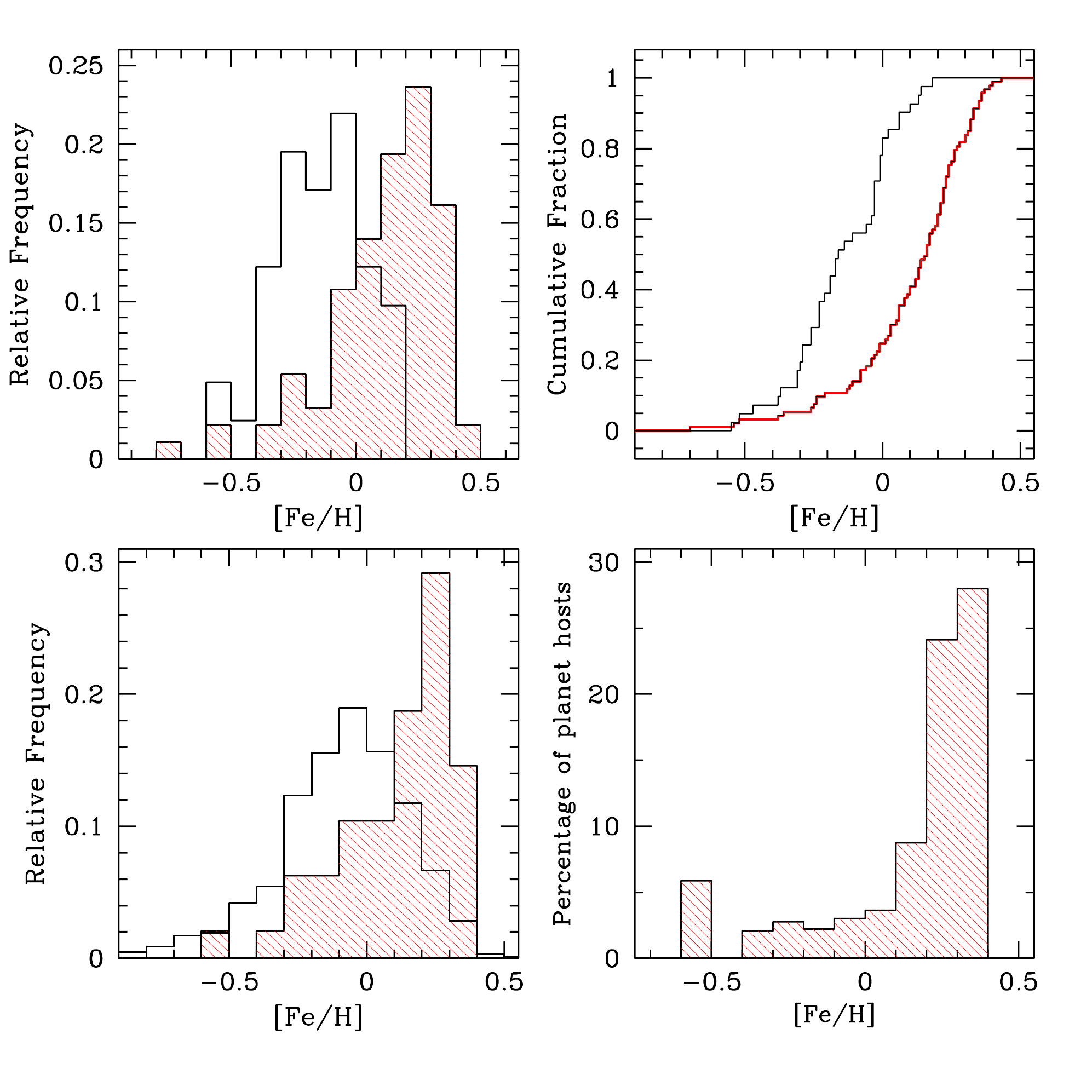}}
\caption{ The planet-metallicity correlation: The iron metallicity distributions for planet host stars (hashed histogram) compared with the distributions of a volume-limited sample of stars (upper-left) and of all the stars in the CORALIE program with at least 5 radial-velocity measurements (lower left). The percentage of planet hosts found amid the stars in the CORALIE sample as a function of stellar metallicity is shown in the lower-right plot. Credit: Santos et al., A$\&$A, 415, 1153, 2004, reproduced with permission \copyright ESO.} \label{santos2004}
\end{figure}

\subsection{Correlations with metallicity}

\subsubsection{Giant planet - metallicity correlation}

Soon after 51 Peg b and 3 other giant planets discoveries, \citet{1997MNRAS.285..403G} analyzed the hosts' high-resolution spectra concluding they had a higher average metallicity than field stars hosting no planets. This led to what is known as the giant planet-metallicity correlation: the higher the star's metallicity, the higher the probability of the star hosting a giant planet, as \fref{santos2004} shows. Although \citet{1997MNRAS.285..403G} proposed that pollution from infall material in the stellar convective envelope was the reason behind this correlation, it has been established that the metallicity of the host reflects the abundance of solids in the primordial cloud that formed the planetary system  \citep[][]{2004A&A...415.1153S,2005ApJ...622.1102F}. The existence of this correlation favors the core-accretion model for planet formation, wherein disks with a higher metallic content, rocky and icy cores can form early enough to allow for runaway accretion to form giant planets before the dissipation of the disks happens, unlike the lower-metallicity disks \citep[][]{1996Icar..124...62P, 2004ApJ...604..388I, 2012A&A...547A.111M}. The exact functionality of the relation is still a matter of study, especially in the metal-poor regime where samples are still small to further constrain the occurrence rates of giants  \citep[][]{2013A&A...557A..70M, 2019Geosc...9..105A, 2021AJ....162...85B}

The original giant planet-metallicity correlation was limited to FGK stars since solar-like stars were the preferred targets of the first RV surveys \citep[][]{2003A&A...398..363S, 2004A&A...415.1153S,2005ApJ...622.1102F}. However, it was soon evident that the correlation did hold for other stars. The few M dwarfs with giant planets discovered by the RV method showed a higher average metallicity when compared with field stars as well  \citep[e.g.][]{2007A&A...474..293B,2012ApJ...748...93R,2012ApJ...747L..38T,2020A&A...644A..68M}. Giant stars with giant planets also showed higher average metallicities when compared with giants without planets, with an overabundance of planets around giant stars with iron metallicity of $\sim$ $-$0.3 dex \citep[][]{ 2015A&A...574A.116R, 2016A&A...590A..38J}.

The spectroscopic characterization of Kepler host stars corroborated that large planets (R$_p$ $>$ 4 R$_\oplus$) are preferentially found around metal-rich stars  \citep[e.g.][]{2012Natur.486..375B,2014Natur.509..593B,2015ApJ...799L..26S}. Results from wide-field ground-based surveys have found that hot Jupiters preferentially orbit metal-rich stars as well, concluding that probably all giants are formed by a similar process, but hot Jupiters have different migration histories \citep[][]{2020MNRAS.491.4481O}.

Unlike the giant planets, rocky and icy planets are not preferentially found around metal-rich stars  \citep[e.g.][]{2008A&A...487..373S,2010ApJ...720.1290G}. However, \citet{2015AJ....149...14W} found a "Universal" planet-metallicity connection for solar-type stars: terrestrial, gas-dwarf, and gas-giant planets occur more frequently in metal-rich stars, but the dependence on metallicity for terrestrial and gas-dwarf planets is lower than for gas giants.

\subsubsection{Planet distance/period - metallicity trend}
 
Several works find trends in stellar metallicity with the orbital period or distance distributions of small and giant planets  \citep[e.g.][]{2013ApJ...763...12B,2015MNRAS.453.1471D}. For example, the occurrence rate of Kepler sub-Neptunes with orbital periods below 10 days as a function of metallicity is three times higher for stars with super-solar metallicity  \citep[][]{2016AJ....152..187M}. Furthermore, planets with masses between 10 M$_{\oplus}$ and 4 M$_{jup}$ orbiting metal-poor stars exhibit longer periods than those orbiting metal-rich stars  \citep[][]{2013A&A...560A..51A}. If the host's metallicity is a proxy of the metal content in the disks, the above results may indicate that planets form farther out from their stars in metal-poor disks or that they form later and do not migrate as inward as planets in metal-rich disks.

\subsection{Correlations with stellar mass}

The mass is the most fundamental property of a star since it determines its whole evolution. As planetary searches extended and samples grew, it was found that stellar mass may play a role in the type of planets that a star can host. Radial velocity surveys soon unraveled the scarcity of giant planets around the most petite stars in their samples. Giant planets with periods of few days should have been easier to discover than rocky planets around red dwarfs, given the favorable mass ratio for detection if they existed. If giant planets were indeed necessary for the evolution of intelligent organisms in a planetary system, as \citet{1994Ap&SS.212...23W}, and \citet{2003AJ....125.2664L} speculated, then a decrease in the incidence of radial velocity giant companions around red dwarfs could have enormous implications for the search of civilizations, since they are the majority of the stars in the Galaxy. \citet{2006ApJ...649..436E} also found a lower frequency of close-in giant planets for M dwarfs from a sample of 90 red dwarfs with RV data from several spectrographs and concluded that their results confirmed theoretical predictions by \citet{2004ApJ...604..388I} and \citet{2004ApJ...612L..73L} about the formation of giants by core-accretion

\citet{2010PASP..122..905J} calculated a functional form of the likelihood of a star (within a mass range from 0.2 to 1.9M $M_\odot$ to harbor a giant planet as a function of mass and metallicity, using more than 1000 stars observed by the California Planet Survey. \citet{2010PASP..122..905J} found that at solar metallicity, the giant planet occurrence rise from 3$\%$ around M dwarfs to 14$\%$ around A stars, and concluded that, if disk masses correlate with stellar mass, this was strong supporting evidence of the core accretion model of planet formation from cool dwarfs to intermediate-mass subgiants.

\begin{figure}[ht]
\centerline{\includegraphics[width=8cm]{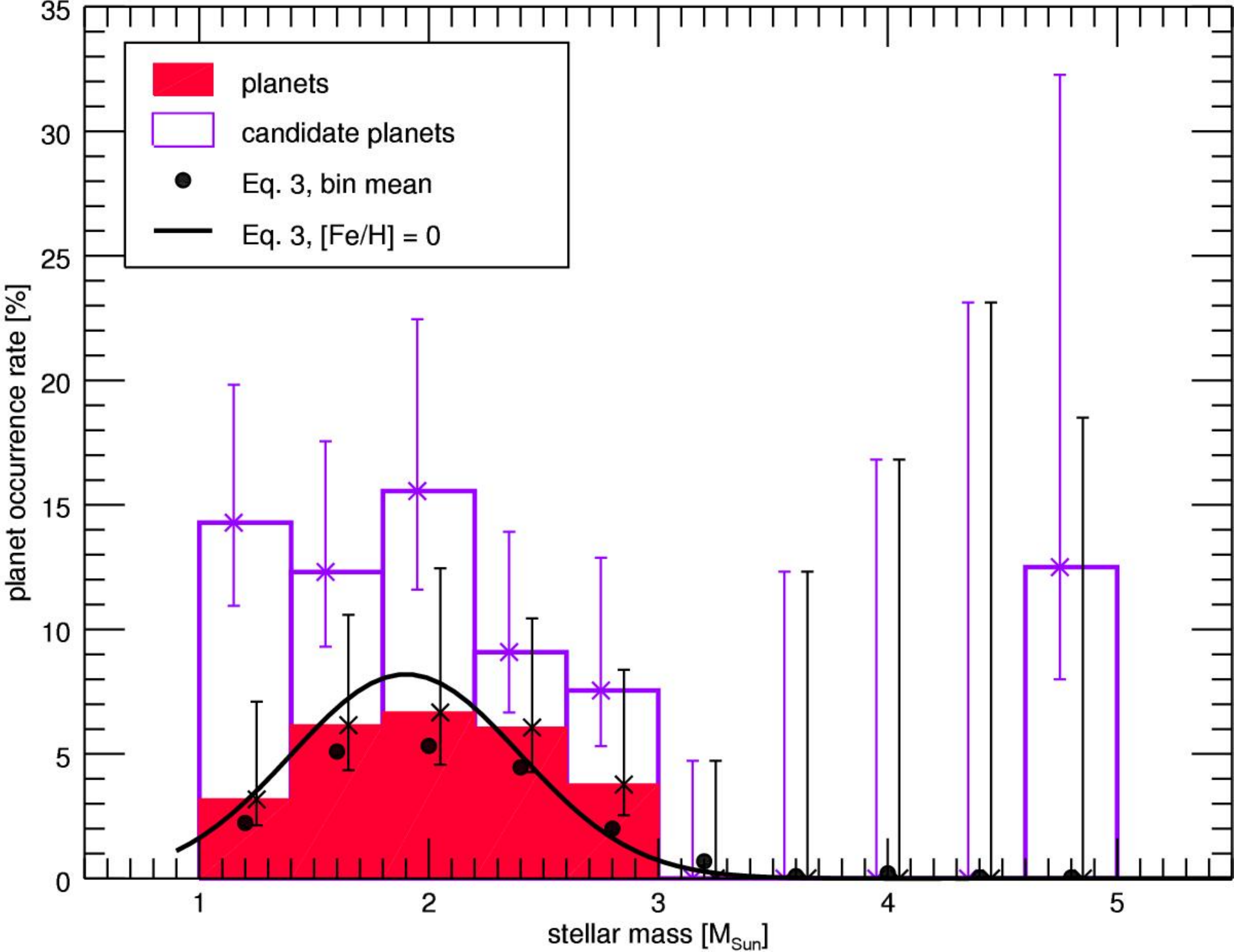}}
\caption{The planet- stellar mass correlation: Planet occurrence rate as a function of stellar mass for giant stars, ignoring the effect of stellar metallicity. The filled histogram shows secure planets, whereas the open histogram includes planet candidates as well. The solid line denotes the best fit to the mass dependence of the giant planet occurrence rate computed for solar metallicity. The black dots correspond to the same model, but the true metallicity distribution within each bin has been taken into account. Credit: Reffert et al., A$\&$A, 574, A116, 2015, reproduced with permission \copyright ESO.} \label{reffert2015}
\end{figure}

Radial velocity surveys also found correlations with mass for giant stars. \citet{2015A&A...574A.116R} and \citet{2016A&A...590A..38J} found that the planet occurrence rate of close-in Jovians increases with stellar mass up to 2 M$_\odot$ for giant stars, but it decreases rapidly and is consistent with zero at $\sim$ 3 M$_\odot$, as shown in \fref{reffert2015}. \citet{2015A&A...574A.116R} propose that stars massive than 2.5 M$_\odot$ may lack giants planets at few AU because their snow line is further out, where gas densities and Kepler velocities are smaller, slowing down growth rates and increasing migration time scales that combine with a shorter lifetime of the protostellar disk, prevent these stars from forming close-in giant planets that would be observable today. 

There is a consensus regarding Neptunes and the sub-Neptune population that they occur more frequently around M dwarfs than FGK stars. Using the Kepler discoveries with periods between 2 to 50 days, \citet{2015ApJ...798..112M}  estimated that for minor planets ($<$ 3 R$_\oplus$), their occurrence rate was higher for M dwarfs, but at larger planet radii  ($>$ 4 R$_\oplus$), the trend reverses, and more giant planets become more common around sun-like stars. These empirical results agree with model calculations, where the core-accretion scenario shows no difficulties in forming low-mass planets around red dwarfs and even predicts more  Neptunes in short orbits around M dwarfs than G dwarfs  \citep[][]{2004ApJ...612L..73L,2004ApJ...604..388I}

Recently, a correlation between protoplanetary disks gaps and stellar mass was found with ALMA observations of 500 young systems.
\citet{2021AJ....162...28V} found that higher mass stars in the sample have relatively more disks with gaps than lower mass stars and that the frequency of the gapped disks matches the observed frequency of giant planets at different stellar masses. If the openings are formed by planets of Neptune mass and above (i.e., giant planets), and they migrate inwards, this result is consistent with the stellar mass - giant planet correlation from exoplanet surveys. The correlation also applies to low-mass stars: disks without gaps are compact, and without giant planets, the dust will drift inwards, providing the necessary conditions for forming more minor, rocky planets with short periods, consistent with the observations of sub-Neptunes mentioned above. Therefore, the mass of host stars can directly relate the exoplanet to its planet-forming environment.

\subsection{Chemical signatures of planet formation}

It is often assumed that a star and its planets form together from the same cloud and have similar compositions. There is a good match in the abundance of refractory elements between the sun and the most primitive and undifferentiated meteorites in our solar system, the CI carbonaceous chondrites \citep{2009ARA&A..47..481A}. Therefore, the atmospheric abundances of refractory elements (such as Mg, Si, Ca, Ni, and Fe) of solar-type stars can be considered a proxy of the composition of the protoplanetary disk. The chemical elements in the disk condensate at different temperatures and therefore condensate at distinct regions of the disk, separating themselves from the gas as dust, while the protostar continues accreting gas. Refractory elements have high condensation temperatures and condensate close to the star-forming rocky planetesimals, while the volatile elements have low condensation temperatures, forming icy planetesimals further away from the star. 

Chemical signatures related to the content of refractory elements in the atmosphere of stars have been searched using high-res, high signal-to-noise spectroscopic data to obtain high precisions in the measurement of element abundances.
\citet{2009ApJ...704L..66M} used a differential method to get precisions of $\sim$ 0.01 dex on solar twins to find that the sun was peculiar, having lower refractory abundances than the average of the solar twins, and proposing that the missing refractories were used to form rocky material in the solar system. \citet{2018ApJ...865...68B} extended the sample of \citet{2009ApJ...704L..66M} up to 79 solar twins, finding that the sun indeed has a deficiency in refractory material relative to more than 80$\%$ of the sample, suggesting that it could be a possible signpost for planetary systems like the solar system in the other refractory poor solar twins. \citet{2019MNRAS.485.4052C} found that the sun also has the lowest lithium abundance compared to the solar twins of the same age, and the most lithium-depleted solar twins were also depleted of refractory elements. The lack of refractory elements or different chemical compositions in planet hosts has been observed in binary systems, such as 16 Cygni \citep[giant planet, ][]{2011ApJ...740...76R,2017A&A...608A.112N, 2019A&A...628A.126M}, {\ensuremath{\zeta}}$^{2}$ Reticuli \citep[debris disk,][]{2016A&A...588A..81S}, WASP-94 \citep[hot Jupiter,][]{2016ApJ...819...19T} and HD 133131 \citep[][]{2016AJ....152..167T}. 

The overabundance of refractory elements and lithium in the atmospheres of stars has been linked to planetary formation, however, as a signpost of planetary accretion or engulfment in stars without detected planets. This is the case of a comoving pair analyzed in \citet{2018ApJ...854..138O}, where it is suggested that one of the components accreted 15 M$_\oplus$ of rocky material after birth to explain the enhancing of refractory elements and lithium found in its atmosphere. In a recent study, \citet{Spina2021} analyzed 107 binary systems of Sun-like stars with similar effective temperatures and surface gravities, concluding that the discrepancies in chemical abundances in the binary systems favored the planet engulfment scenario and estimating that it occurs in about a quarter of all sun-like stars. Pollution or engulfment of differentiated material has also been observed in the atmospheres of white dwarfs  \citep[e.g][]{2010ApJ...722..725Z,2021MNRAS.503.1877B}

Recently, a correlation between the compositions of rocky exoplanets and their host stars was found. By concentrating only on planets with masses below 10 M$_\oplus$ but avoiding mini-Neptunes, \citet{2021Sci...374..330A} found that the iron content of rocky planets (inferred from their estimated density) correlates with the iron content of the star, which reflects the iron content of the protoplanetary disk. However, it is not a one-to-one correlation as was expected. Instead, the planets are more enhanced in iron than their host stars. The reason behind this is still unknown, but \citet{2021Sci...374..330A} suggest that it can be related to rocky lines (condensation/sublimation lines of refractory materials) in the disk where the fraction of iron can be enhanced, as described in \citet{2020ApJ...901...97A}.

\section{Conclusions}

The planet discovery techniques and methods prove different types of stars and distinct areas of our neighborhood and populations of the Galaxy. For most of the planetary systems known to date, all that we can see are the stars; hence, we need to know the stars in detail to obtain the exact bulk properties of the exoplanets and their orbital parameters, calculate the occurrence rates, and infer how the formation and evolution of the planetary systems occurred.  

The stellar mass and metallicity are quantities showing connections with specific types of exoplanets in the stars surveyed. These two quantities are believed to depict the amount of material and the composition of the cloud that formed the planetary system. Knowing them in detail makes it possible to assume specific disk characteristics, figure out how relevant the initial conditions are for the formation of planets, and find possible explanations for the planetary systems that don't follow the expected trends. 

Regarding the metallicity of the stars, hosts of giant planets on average have higher metallicities than field stars (i.e., are metal-rich stars), a result known as the giant planet-metallicity correlation that supports the core accretion scenario for planet formation. On the other hand, the hosts of only small planets exhibit a wide range of metallicities. Furthermore, if we consider the orbital period of the planets and metallicity, we find that Neptunes and super-Earths with short periods are found preferentially around metal-rich stars, while sub-Neptunes to giants with longer periods are found preferentially around metal-poor stars.

Regarding the mass of the stars, the current samples of exoplanetary systems show that the occurrence rate of giant planets increases as stellar mass increases, however only up to 2 M$_\odot$, where it decreases as stellar mass increases. The occurrence rate of sub-Neptunes increases as stellar mass decreases.

Regarding chemical signatures of planet formation, works have found the sun peculiar compared with samples of solar twins, being poor in refractory elements and lithium. The overabundance of refractory elements and lithium in the atmospheres of stars has been proposed as a signpost of planetary engulfment in stars without detected planets. 

The star-planet connection is an area in constant review, where newer discoveries allow the recognition and constraining of the physical processes involved in the formation and evolution of planetary systems. Therefore, it should not be surprising that the relationships presented in this chapter become increasingly specific concerning the characteristics and detection method of the system or exoplanet in question or modified to include new observations, which allow new connections to be found.

\bibliographystyle{aa}
\bibliography{brojasayala}


\end{document}